\documentstyle[twoside,fleqn,espcrc2,epsf]{article}

\newcommand{\eq}{\begin{equation}}
\newcommand{\en}{\end{equation}}
\newcommand{\eqa}{\begin{eqnarray}}
\newcommand{\ena}{\end{eqnarray}}

\newcommand{\AmS}{{\protect\the\textfont2
  A\kern-.1667em\lower.5ex\hbox{M}\kern-.125emS}}

\title{Glueballs and string breaking from full QCD\thanks{Presented by
    Gunnar~Bali.}}

\author{SESAM and T$\chi$L collaborations: G.S.~Bali$^{\rm a}$,
        N.~Eicker$^{\rm b}$, L.~Giusti$^{\rm c}$,
        U.~Gl\"assner$^{\rm d}$,
        S.~Guesken$^{\rm d}$, H.~Hoeber$^{\rm d}$,
        P.~Lacock$^{\rm b}$, T.~Lippert$^{\rm b}$,
        G.~Martinelli$^{\rm c}$, F.~Rapuano$^{\rm c}$,
        G.~Ritzenh\"ofer$^{\rm e}$, K.~Schilling$^{\rm b,d}$, 
        G.~Siegert$^{\rm b}$, A.~Spitz$^{\rm b}$,
        P.~Ueberholz$^{\rm d}$ and J.~Viehoff$^{\rm d}$\\[8pt]
{\rm $^a$}Department of Physics, The University, Highfield,
           Southampton SO17 1BJ, UK\\[8pt]       
{\rm $^b$}HLRZ c/o Forschungszentrum J\"ulich, D-52425 J\"ulich
          and DESY, D-22603 Hamburg, Germany\\[8pt]  
{\rm $^c$}Dip.\ di Fisica, Univ.\ ``La Sapienza'' and
INFN, Sezione di Roma, P'lle A.\ Moro, I-00185 Rome, Italy\\[8pt]
{\rm $^d$}Physics Department, University of Wuppertal, D-42097
           Wuppertal, Germany\\[8pt]
{\rm $^e$}MIT, Center for Theoretical Physics, Cambridge,
Massachusetts 02139, USA}
      
\begin{document}

\begin{abstract}
We present results on the static potential, and torelon and glueball masses
from simulations of QCD with two flavours of dynamical Wilson fermions
on $16^3\times 32$ and $24^3\times 40$ lattices at $\beta=5.6$.
\end{abstract}

\maketitle

\section{SIMULATION}
\noindent
The present simulations have been performed at various $\kappa$ values and
lattice volumes at $\beta=5.6$ (Table~\ref{tab1}).
The effective lattice resolution ranges
from $a^{-1}\approx 2.0$~GeV ($\kappa=0.156$) down to
$a^{-1}\approx 2.5$~GeV ($\kappa=0.158$) while the
ratio $m_{\pi}/m_{\rho}$ varies from 0.83 to 0.55,
corresponding to sea quarks that are slightly heavier than the strange
quark and of about one quarter of its mass, respectively.
In addition, results from quenched reference
simulations at $\beta= 6.0$ ($a^{-1}\approx
2.1$~GeV) and $\beta= 6.2$ ($a^{-1}\approx 3.1$~GeV) are presented.

\begin{table}[hbt]
\vskip -.6cm
\setlength{\tabcolsep}{0.7pc}
\newlength{\digitwidth} \settowidth{\digitwidth}{\rm 0}
\catcode`?=\active \def?{\kern\digitwidth}
\caption{Simulation parameters.}
\label{tab1}
\begin{tabular}{lrrrr}
\hline
$\kappa$&$V$&$r_0a^{-1}$&$n_{\mbox{\scriptsize glue}}$&$n_{\mbox{\scriptsize pot}}$\\
\hline
0.1560&$16^332$&5.11(3)&2129&236\\
0.1565&$16^332$&5.28(5)& --- &323\\
0.1570&$16^332$&5.46(5)&2039&240\\
0.1575&$16^332$&5.98(7)&2272&270\\
0.1575&$24^340$&5.93(4)&1243&122\\
0.1580&$24^340$&6.33(7)&743&95\\\hline
$\beta=6.0$&$16^332$&5.33(3)& --- &570\\
$\beta=6.2$&$32^4$&7.29(4)& --- &116\\
\hline
\end{tabular}
\vskip -.7cm
\end{table}

On each of the small volumes, about 5000 thermalized trajectories have been
generated. One half of this amount has been achieved on the $24^3\times 40$
lattice at $\kappa=0.1575$ while the $\kappa=0.158$ run has not been
completed yet. Glueball and torelon measurements
($n_{\mbox{\scriptsize glue}}$)
have been taken every 2 trajectories while smeared Wilson loops
($n_{\mbox{\scriptsize pot}}$)
were measured every 20 trajectories (16 at $\kappa=0.1565$).
Prior to statistical analysis, all data have been binned
into blocks whose extent was large compared to the relevant
autocorrelation time (see Ref.~\cite{lippert}).

\section{THE STATIC POTENTIAL}
\noindent
An increase of the strength of the
Coulomb-like attractive force between static sources
at small separations, in
respect to
the quenched limit, has been observed previously~\cite{wachter};
the QCD coupling is running
slower with the energy scale, once sea quarks are switched on.
We confirm these findings with increased statistical accuracy
and full control over systematic uncertainties
(see Ref.~\cite{guesken}).  
Moreover, in extrapolating the Sommer scale
$r_0^{-1}$ as a polynomial in the quark mass
to the physical limit, we find a $\beta$ shift,
$\beta_{n_f=0}
-\beta_{n_f=2}=0.57(2)$. This corresponds to an
increase in the coupling of about 10\% which is close
to the na\"\i{}ve perturbative ratio $33/(33-2n_f)$.

With dynamical fermions, the static meson can decay into two static-light
mesons. Ignoring meson-meson interactions,
we expect the QCD string to ``break''
as soon as the
potential exceeds twice the static-light mass,
i.e.\ at about 1.5 fm. Neglecting quark mass effects
on the dynamics of the binding problem
(which is a reasonable assumption once this mass is small
compared to a typical binding energy of 500 MeV),
the string breaking distance should be shifted
by an amount $\Delta r\approx 2\Delta m/\sigma$ when
changing the quark mass by $\Delta m$. $\sigma = Ka^{-2}$
denotes the {\em effective} string tension.
From these considerations, we find $\Delta r < 0.25$~fm for
$\kappa \geq 0.156$.
However, up to a separation
of 2~fm no indications of a flattening of the potential are found.
We suspect that this
is due to a bad overlap of the Wilson loop with
the two-meson state.

\begin{figure}[htb]
\leavevmode
\vskip -.75cm
\epsfxsize=7.5cm\epsfbox{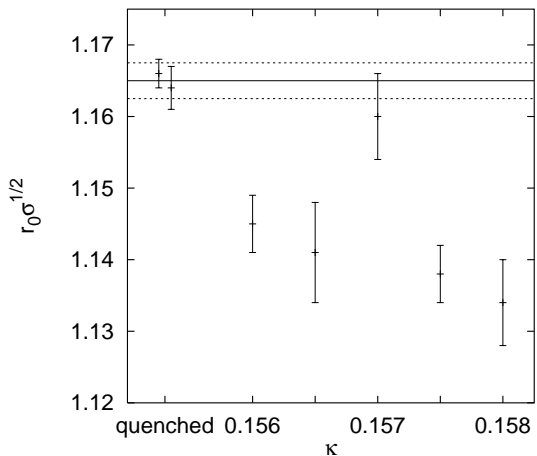}
\vskip -1.2cm
\caption{The effective string tension $\sigma$ in units of the
Sommer scale $r_0$.\vskip -.6cm}
\label{fig1}
\end{figure}

Regg\`e trajectories suggest values of 420 to 440~MeV for the string tension.
Fits of potential models to the bottomonium spectrum yield similar
values. However, such parametrizations face difficulties in setting
a reliable upper bound on
$\sigma$; effects of an increased slope of the potential at large distance
can be absorbed by decreasing the strength of the short range
interaction. This is due to the fact that the spectrum is rather insensitive
towards the shape of the potential outside of a region $0.2$~fm~$<r<1$~fm.
In contrast, the scale $r_0$ which is defined through
the interquark force at intermediate distance can be constrained
to $r_0^{-1}\approx 395$~MeV rather accurately.
In Fig.~\ref{fig1}, we display the value $r_0\sqrt{\sigma}$ against
$\kappa$. While a quenched value
$\sqrt{\sigma}\approx 460$~MeV appears to be reasonable, with 
two sea quarks
this is reduced to about 450~MeV
such that for three active sea quarks a picture,
consistent with the slope
of Regg\`e trajectories, is likely to emerge.

\section{TORELON STATES}
\noindent
Torelons are flux tubes encircling the periodic spatial lattice boundaries.
We restrict ourselves to torelons with winding number one.
As a consequence of the
center group symmetry all such torelons are
degenerate in the quenched case. Moreover, the mass of
such a state is expected to equal
$L_Sa\sigma$ (up to tiny finite size corrections),
where $L_Sa$ denotes the spatial extent of the lattice.

\begin{figure}[htb]
\leavevmode
\vskip -1cm
\epsfxsize=7.5cm\epsfbox{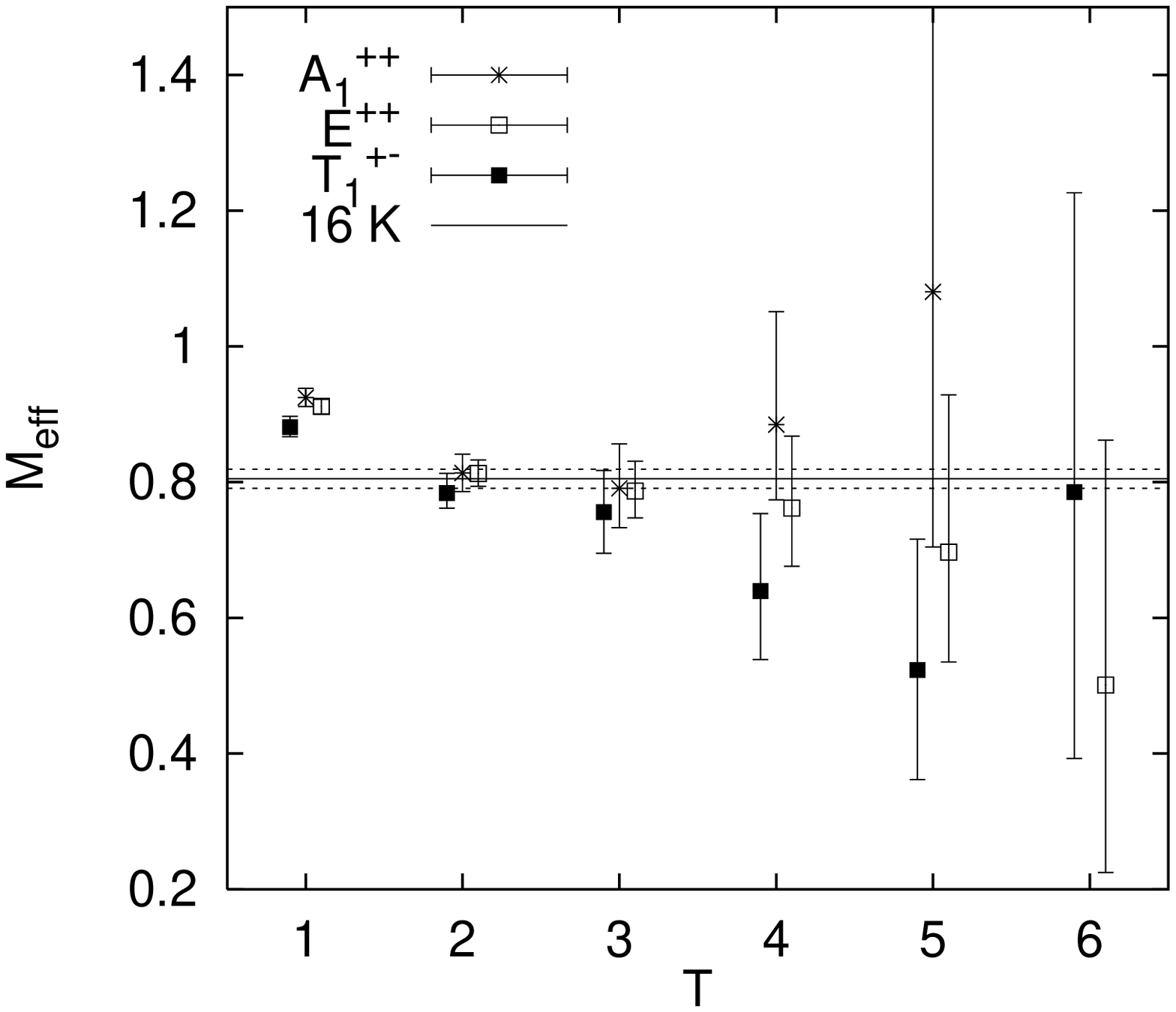}
\vskip -1.2cm
\caption{Effective torelon masses at $\kappa = 0.156$.\vskip -.8cm}
\label{fig2}
\end{figure}

With sea quarks, the $Z_3$ symmetry is only approximate
and various torelon states, that can be classified in accord with
irreducible representations of the cubic symmetry group, can obtain
different masses. Moreover, mixing (or decays) between torelon states
and isoscalar mesons might occur.
The $T_1^{+-}$ torelon is accompanied by the lightest isoscalar,
the continuum $J=1$ $h_1(1190)$ meson.

\begin{figure}[htb]
\leavevmode
\vskip -.15cm
\epsfxsize=7.5cm\epsfbox{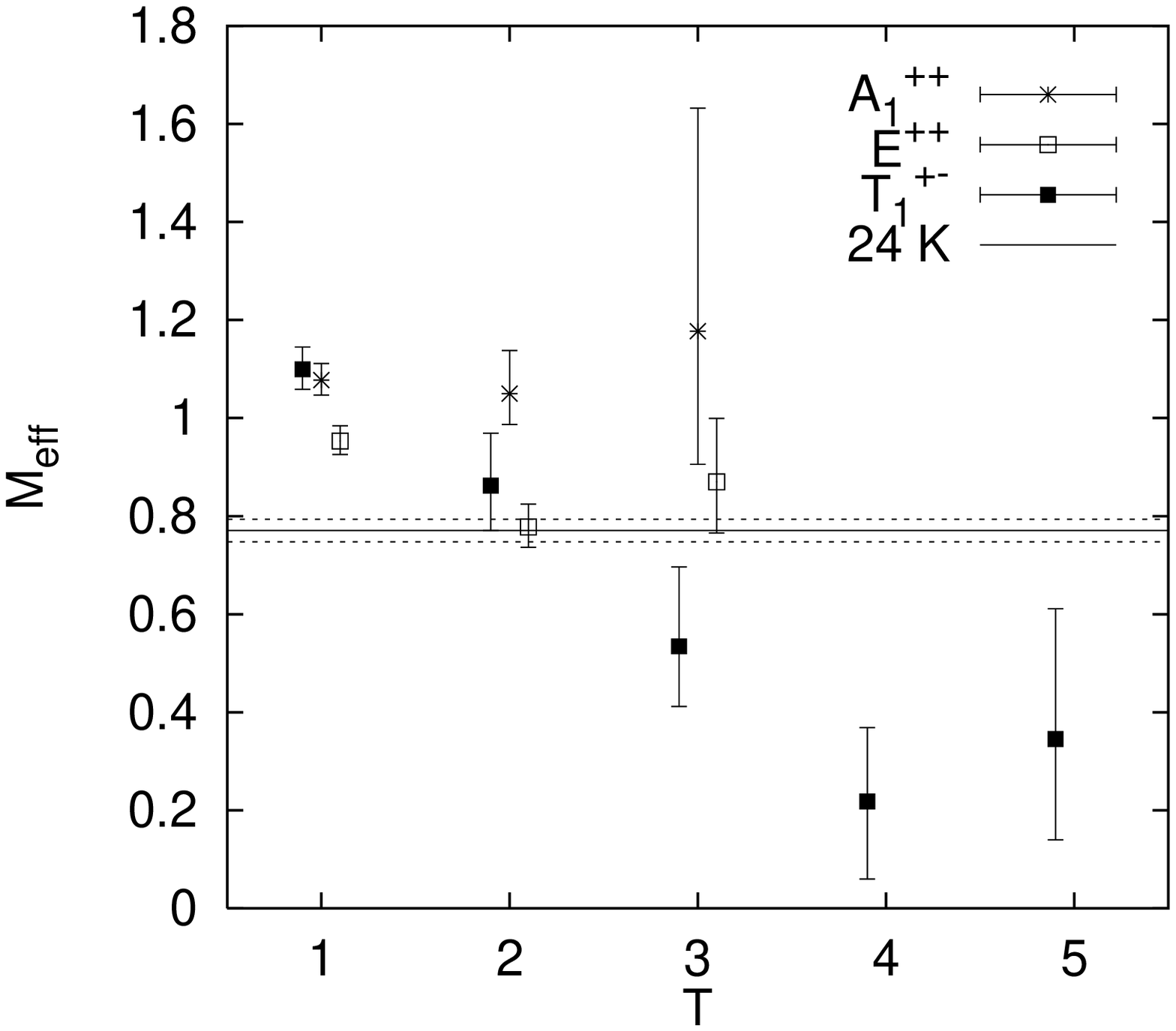}
\vskip -1.2cm
\caption{Effective torelon masses at $\kappa = 0.158$.\vskip -0.7cm}
\label{fig3}
\end{figure}

In Fig.~\ref{fig2} we display effective masses of three different torelon
states at $\kappa=0.156$. Note that all effective masses are strict
upper limits on the real mass value. All states are degenerate and in
agreement with the expectation $L_S a\sigma$. The same holds true for
$\kappa=0.157$. However, at $\kappa=0.1575$ and $\kappa=0.158$ we see
indications of certain torelon states becoming lighter than expected.
This situation is depicted in Fig.~\ref{fig3} for $\kappa=0.158$,
where
the $T_1^{+-}$ state falls below $24a\sigma$. Signals of the
other states disappear into noise.

\section{GLUEBALLS}
\begin{figure}[htb]
\leavevmode
\vskip -.15cm
\epsfxsize=7.5cm\epsfbox{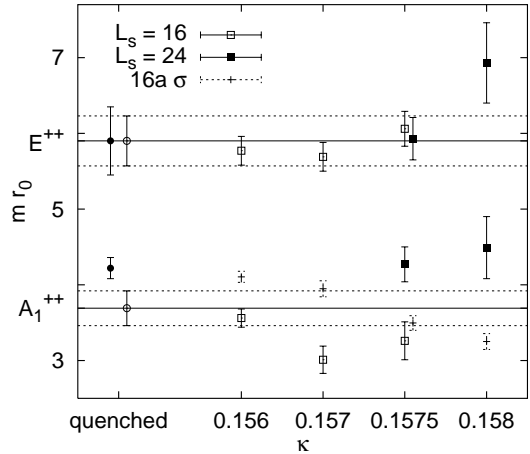}
\vskip -1.2cm
\caption{Glueball masses. Solid and open circles denote quenched
results, extrapolated to the continuum, and at $\beta=6.0$,
respectively.\vskip -.7cm}
\label{fig4}
\end{figure}

\noindent At sufficiently light sea quark masses we expect the
$0^{++}$ glueball to mix with the two
mesonic $I=0$ states of the $L=1$
nonet. Moreover, decays into two $\pi$'s become possible.
Results on the scalar ($A_1^{++}$) and tensor
($E^{++}$) glueballs are displayed in
Table~\ref{tab2} and Fig.~\ref{fig4}.
No indications of
the mentioned effects are found.

\begin{table}[hbt]
\vskip -.6cm
\setlength{\tabcolsep}{0.8pc}
\catcode`?=\active \def?{\kern\digitwidth}
\caption{Glueball masses.}
\label{tab2}
\begin{tabular}{lrrr}
\hline
$\kappa$&$L_S$&$m_{0^{++}}r_0$&$m_{2^{++}}/m_{0^{++}}$\\
\hline
0.1560&16&3.56(12)&1.62(07)\\
0.1570&16&3.01(18)&1.89(13)\\
0.1575&16&3.26(25)&1.86(13)\\
0.1575&24&4.27(23)&1.46(28)\\
0.1580&24&4.49(41)&1.54(21)\\\hline
$\beta=6.0$&20&3.69(23)&1.67(15)\\
$\beta=\infty$&---&4.22(14)&1.40(15)\\
\hline
\end{tabular}
\vskip -.7cm
\end{table}

In quenched simulations, finite size effects
on the scalar glueball are small~\cite{michael2} as long as
$m_{0^{++}}<2m_t\approx 2L_Sa\sigma$
where $m_t$ denotes the torelon mass. 
When including sea quarks, the glueball
might break up and decay into a single torelon~\cite{michael3}
through an intermediate mesonic
state. The expected torelon masses $L_Sa\sigma$ are
included into Fig.~\ref{fig4} for $L_S=16$
(dashed error bars). Contrary to the
quenched case, we indeed find significant FSE at $\kappa=0.1575$ (and
probably $\kappa=0.157$).
In general, the large volume data is consistent
with quenched results. The $0^{++}$ mass appears to
exceed the $\beta=6.0$ reference value but
agrees with the continuum extrapolated result.

\section*{ACKNOWLEDGEMENTS}
\noindent
We acknowledge support by the DFG (grants Schi 257/1-4 and Schi
257/3-2), the EU (contracts SC1*-CT91-0642, CHRX-CT92-0051
and CHBG-CT94-0665) and PPARC (grant GR/K55738).


\begin{thebibliography}{9}
\bibitem{lippert} T$\chi$L collaboration: T.~Lippert et al.,
these proceedings.

\bibitem{wachter} SESAM collaboration: U.~Gl\"assner et al.,
Phys.~Lett.~B383 (1996) 98.

\bibitem{guesken} S.~Guesken, these proceedings.





\bibitem{michael2} C.~Michael and M.~Teper, Nucl.~Phys.~B314 (1989) 347.

\bibitem{michael3} J.~Kripfganz and C.~Michael, Nucl.\ Phys.\ B314 (1989) 25.
\end{thebibliography}
\end{document}